\documentclass[journal]{IEEEtran}
\usepackage{cite}

\usepackage{subfigure}
\ifCLASSINFOpdf
\usepackage[pdftex]{graphicx}
  \graphicspath{{../pdf/}{../jpeg/}}
  \DeclareGraphicsExtensions{.pdf,.jpeg,.png}
\else
  \usepackage[dvips]{graphicx}
  \graphicspath{{../eps/}}
  \DeclareGraphicsExtensions{.eps}
\fi  
\usepackage{amsmath}
\usepackage{cases}
\usepackage{stfloats}
\usepackage{amsfonts}
\usepackage{subeqnarray}
\usepackage{longtable}
\usepackage{supertabular}
\usepackage{setspace}
\usepackage{multirow}

\usepackage[ruled,linesnumbered]{algorithm2e}
\makeatletter
\newcommand{\nosemic}{\renewcommand{\@endalgocfline}{\relax}}
\newcommand{\dosemic}{\renewcommand{\@endalgocfline}{\algocf@endline}}
\let\oldnl\nl
\newcommand{\nonl}{\renewcommand{\nl}{\let\nl\oldnl}}
\makeatother
\usepackage[export]{adjustbox} 
\usepackage{booktabs}
\usepackage{setspace}
\usepackage{xcolor}
\usepackage{mathrsfs}
\usepackage{amsmath}
\usepackage{array}
\usepackage{amssymb}
\usepackage{amsthm}
\usepackage{microtype}
\usepackage{url}
\usepackage{amsfonts,amssymb}
\usepackage{dsfont}
\usepackage{mathtools}
\usepackage{xcolor,colortbl}
\usepackage{colortbl}

\definecolor{Gray}{gray}{0.85}
\definecolor{Whitecolor}{rgb}{1,1,1}

\newcolumntype{a}{>{\columncolor{Gray}}c}
\newcolumntype{b}{>{\columncolor{white}}c}

\allowdisplaybreaks
\begin{document}
\setstretch{0.93}
\title{\textls[-25]{Control-aware Probabilistic Load Flow for Transmission Systems: An Analytical Method}}
\author{Mengshuo~Jia,~\IEEEmembership{Member,~IEEE,}  
Qianni~Cao,~\IEEEmembership{Student~Member,~IEEE}, 
Chen~Shen,~\IEEEmembership{Senior~Member,~IEEE}, 
and
Gabriela~Hug,~\IEEEmembership{Senior~Member,~IEEE}

}
        

\maketitle

\begin{abstract}
	Probabilistic load flow (PLF) calculation, as a fundamental tool to analyze transmission system behavior, has been studied for decades. Despite a variety of available methods, existing PLF approaches rarely take system control into account. However, system control, as an automatic buffer between the fluctuations in random variables and the variations in system states, has a significant impact on the final PLF result. To consider control actions' influence, this paper proposes the first analytical PLF method for the transmission grid that takes into account primary and secondary frequency controls. This method is based on a high-precision linear power flow model, whose precision is even further improved in this paper by an original correction approach. This paper also proves that if the joint probability distribution (JPD) of random variables is expressed by a Gaussian mixture model (GMM), then the JPD of system states (e.g., nodal voltages) is an infinite GMM. By leveraging this proposition, the proposed method can generate the joint PLF of the whole system, is applicable to random variables obeying any distributions, and is capable of capturing their correlation. The high accuracy and satisfactory efficiency of this method are verified on test cases scaling from 14 to 1354 buses. 

\end{abstract}
\IEEEpeerreviewmaketitle

\section{Introduction}

\IEEEPARstart{V}{ariable} renewables continually inject uncertainties into the transmission system, leading to random fluctuations of system states (i.e. nodal voltages and branch flows). Probabilistic load flow (PLF) calculations can effectively capture the uncertainties in these states and evaluate the potential operational risk \cite{PRUSTY20171286}, thereby not only becoming a fundamental tool to study the behavior of the system \cite{8462798}, but also being applicable to many areas, including daily system operation \cite{5558710,usaola2009probabilistic}, short-term/long-term planning \cite{5682076,4523658}, etc.



Since the first PLF method was proposed in 1974 \cite{4075419}, PLF approaches developed rapidly and pluralistically, resulting in three main categories to date: simulation methods (also known as Monte Carlo methods or scenario methods) \cite{XIAO2018677,liu2016probabilistic}, approximated methods \cite{4349110, 6192342}, and analytical methods \cite{VILLANUEVA20141,7574307}. The core idea of simulation methods is to feed the power flow model with hundreds of thousands of scenarios, yielding numerous power flow results and accordingly summarize the PLF from those results. Scenario reduction techniques \cite{8845655,ruan2020constructing,ruan2020assessment} can be applied, but this generally reduces the accuracy. Approximated methods also need power flow calculation results to estimate the PLF, yet the scenarios used in these methods are specifically chosen according to the moment information of the random variables. As a result, only a few power flow computations are required by approximated methods. Instead of summarizing the PLF from power flow results, analytical methods directly map the uncertainty characteristics (e.g., moments or probability distributions) of random variables to that of system states through simplified power flow models, thereby obtaining the PLF results. Overall, it is broadly agreed that the simulation approach, especially the AC-power-flow-based Monte Carlo simulation with simple random sampling, is the most accurate among the three categories \cite{8049316}, whereas the approximated and analytical methods both realize a compromise between accuracy and computational burden.

However, the majority of existing PLF methods have not taken control actions of the system into account \cite{XIAO2018677,liu2016probabilistic,4349110,6192342,VILLANUEVA20141,7574307}, leading to deviations in the actual and modeled probabilistic flows \cite{zhu2014probabilistic}. In transmission systems, one of the main controls is the frequency regulation, which will either be triggered locally within seconds (i.e. primary control) or centrally at a slower timescale (i.e. secondary control) when there is a power imbalance. Given the fluctuations of renewable energies, the frequency regulation actions may lead the system to another equilibrium point \cite{7764193}, which changes the possible values of all system states. Hence, the PLF of the transmission system is not solely determined by the characteristic of random variables --- the frequency regulation also has a significant impact on the PLF result. It could be argued that the results obtained when ignoring controls reflect the PLF right before any remedial actions are taken \cite{4523658}. But it should be noted that the primary control will be activated automatically after the disturbance \cite{Swissgrid2019}. That is to say, it is likely more reasonable to indeed account for the control actions instead of considering a state that is only present during a short transitional period. It can also be argued that controls can be neglected for PLF calculations due to the assumption that the imbalanced power is completely absorbed by the swing bus \cite{zhu2014probabilistic,7764193}. However, the power imbalance is compensated either by all the conventional generators equipped with a speed governor (primary control) or by all the units participating in the automatic generation control (secondary control). Hence, the assumption above might not hold in practice. 

Although control actions should be taken into account when computing the PLF for transmission systems, there is hardly any literature covering this topic. To the best of our knowledge, only two simulation methods \cite{zhu2014probabilistic,8845655} and one approximated method \cite{7764193} have been proposed to calculate the control-aware PLF for transmission networks. Among them, both \cite{zhu2014probabilistic} and \cite{8845655} are Monte-Carlo-simulation-based. In order to improve the computational efficiency, \cite{zhu2014probabilistic} adopts a piece-wise linearized power flow model while \cite{8845655} leverages the Latin hypercube sampling technique, thereby inevitably harming the PLF accuracy. Additionally, \cite{zhu2014probabilistic} ignores the correlation between random variables (i.e., the outputs of wind farms). In contrast, \cite{8845655} adopts Gaussian Copula to establish the correlation between random variables. Yet, Gaussian Copula requires multiple integral calculations, thus rendering the simulation method in \cite{8845655} computationally expensive. As the only control-aware approach based on approximated methods, \cite{7764193} employs the $2m+1$ point estimation algorithm. Though being efficient, the $2m+1$ method can only generate partial PLF results, i.e. the moments of system states instead of their probability distributions, making the system operators fail to answer probability-related questions, such as what the probability is that the active power flow of a specific line may exceed its thermal limit. To summarize, there exists no analytical control-aware PLF method for transmission systems, and existing approaches still suffer from the issues above. 


This motivates us to develop an analytical PLF algorithm, and at the same time to tackle the aforementioned deficiencies of other types of methods. Specifically, except for meeting the basic requirements, e.g. being analytical, accurate, efficient, and capable of dealing with the primary and secondary frequency controls, we strive to take it a step further by adding more superior features to our approach, namely (1) the proposed method should not require knowledge of or assume a specific probability distribution of the random variable; (2) it should be straightforward for the proposed method to characterize the correlation between random variables, instead of using additional time-consuming techniques; (3) the proposed method should be able to derive the joint PLF result, i.e. the joint probability distribution (JPD) of all system states, which not only contains the marginal probability distributions of system states but also records the relationship between them, thus being far more informative than only knowing the moments. With the joint PLF result, system operators can answer more complex but necessary questions, e.g. what the probability is that no system states violate limits at the same time.

To realize the desired analytical joint PLF method, this paper adopts two basic models: (1) the state-independent voltage-angle decoupled linearized power flow (DLPF) model \cite{7782382} and (2) a Gaussian mixture model (GMM), which simulates the JPDs of the input and output of the PLF calculation. Choosing the DLPF model is due to its high precision and enables the derivation of a piece wise linear relationship between the system states and random variables. The reason for using GMM as the basic probability model is that GMM is able to fit the complex characteristics of arbitrary probability distributions, including skewness, multi-peak, and heavy-tail \cite{8404026}, making GMM a universal distribution model and adaptive to any random variables (wind power, PV, etc.). Also, GMM's covariance can easily capture the correlation between random variables \cite{WANG2018771}. Based on these two models, the problem of computing the joint PLF of transmission systems reduces to the question of how to analytically and accurately translate the GMM of random variables into JPDs of system states when considering frequency controls. To answer this question, we present the following contributions in this paper:
\begin{itemize}
	\item Prove that if the JPD of random variables is expressed by a GMM, then the JPD of the random variables' piece-wise linear function is an infinite GMM.
	\item Propose two approaches to analytically and accurately map the JPD of random variables to the JPD of their piece-wise linear transformation.
	\item Design a polynomial-fitting-based correction method to further improve the precision of the DLPF model.
	\item Propose an analytical, control-aware joint PLF algorithm for transmission systems with high accuracy and satisfactory efficiency. 
\end{itemize}

The remainder of this paper is organized as follows: in Section II, we formulate a power flow model that incorporates the frequency regulation model. Section III presents and solves two critical challenges when developing the analytical joint PLF method, namely how to analytically map the JPD of random variables to the JPD of system states, and how to improve the accuracy of the linear power flow model adopted. The complete method is then proposed in Section IV with detailed discussions on its characteristics. In Section V, various evaluations and comparisons are carried out. Finally, Section VI concludes this paper with outlooks.

\vspace{-0.3cm}
\section{Control-aware Power Flow Model}
In this section, the frequency regulation model, including both primary and secondary controls, are first revisited. Then, we integrate these frequency controls into the power flow model, resulting in a control-aware power flow model, which is the foundation for calculating the joint PLF of the whole system.
\vspace{-0.4cm}
\subsection{Frequency Regulation Model}

Let $\mathcal{S}$ represent the set of $PV$ and $PQ$ nodes, and let $\mathcal{L}$ represent the set of $PQ$ nodes. We also define $N = \rm{card}(\mathcal{S})$ and $M = \rm{card}(\mathcal{L})$, where $\rm{card}(\cdot)$ is the cardinality function. We assume that each $PV$ or $PQ$ bus could have a conventional generator, a renewable generator, and a constant load connected. Accordingly, the active power injection at bus $n$ and the reactive power injection at bus $m$ can be respectively described by
\begin{align}
	P_n & = P_{g,n} + P_{w,n} - P_{d,n}\quad n \in \mathcal{S} \notag \\
	Q_m & = Q_{g,m} + Q_{w,m} - Q_{d,m}\quad m \in \mathcal{L} \notag
\end{align}
where subscript $g$ denotes the conventional generator, subscript $w$ represents the renewable generator, and subscript $d$ refers to the load. Besides, $P$ and $Q$ stand for the active and reactive power. In case of the conventional generators, $P_{g,n}$ corresponds to the scheduled generation without the contribution to frequency control. If no conventional generator is connected to bus $n$, then $P_{g,n}=0$. This paper only considers the uncertainties of renewable generators' outputs, i.e. $P_{w,n}$ and $Q_{w,m}$, as load uncertainty is relatively small compared to renewables \cite{wang2016risk}.

Given that $P_{w,n}$ is a random variable, the overall system may face an active power imbalance, which is given by 
\begin{align}
	P_{\Delta} = \sum\nolimits_{n =1}^{N}(P_{g,n} + P_{w,n} - P_{d,n})\quad  n \in \mathcal{S}  \notag
\end{align}
where the power loss is ignored. The amount of $P_{\Delta}$ is indicative of the resulting frequency deviation and determines which control action needs to be taken. Specifically: 1) In order to avoid frequent actions of speed governors in generators, operators manually set dead zones for generators in the primary frequency control. I.e., when $P_{\Delta}$, or say, the frequency deviation, is less than the dead zone threshold, only frequency-dependent loads react to reduce the deviation; otherwise, both loads and generators respond to control the frequency. 2) When the primary frequency control is able to keep the frequency change within the allowable range, there is no need to activate the secondary frequency control. To this end, operators will set an AGC threshold for AGC units --- when $P_{\Delta}$, or say, the frequency deviation, is less than the AGC threshold, only the primary frequency control acts; otherwise, the secondary frequency control is activated. 

Given the above situations, the relationship between $P_{\Delta}$ and the frequency deviation $f_{\Delta}$ can be expressed by 
\begin{equation}
	f_{\Delta} = \left\{
		\begin{split}
		& P_{\Delta}/K_D &  \quad |P_{\Delta}| \leq K_{D}f_{D} &  \\
		& P_{\Delta}/K_{U} & \quad K_{D}f_{D} < |P_{\Delta}| & \\
		\end{split} \right. \label{eq:f_and_p}
\end{equation}
Note that $f_D$ is the average frequency threshold of all generators' dead zones, while $K_D$ and $K_U$ are defined as:
\begin{align}
	K_D & = \sum\nolimits_{n=1}^N K_{d,n} \quad n \in \mathcal{S} \notag \\
	K_U &  = \sum\nolimits_{n=1}^N K_{g,n} + K_{d,n} \quad n \in \mathcal{S}  \notag 
\end{align}
where $K_{d,n}$ represents the load frequency characteristic coefficient at bus $n$ and $K_{g,n}$ denotes the governor response coefficient of the conventional generator at bus $n$ \cite{7764193}. If no load or generator is connected to bus $n$, $K_{d,n}=0$ or $K_{g,n}=0$. 


Based on \eqref{eq:f_and_p} and the piece-wise characteristic of the frequency regulation aforementioned, the frequency regulation model, i.e. the amount of the active power regulation at bus $n$, can be further detailed as follows:
\begin{equation}
	P_{\Delta,n} = \left\{
		\begin{split}
		& -K_{d,n}P_{\Delta}/K_D  & 0 <|P_{\Delta}| \leq K_{D}f_{D} \\
		& -K_{u,n}P_{\Delta}/K_U   &   K_{D}f_{D} < |P_{\Delta}| \leq K_{U}f_{A}\\
		& -H_{g,n}P_{\Delta}/H_G   &  K_{U}f_{A} < |P_{\Delta}| \leq P_{\Delta, max} \\
		\end{split} \right. \label{eq:FR_model}
\end{equation}
where $K_{u,n} = K_{g,n} + K_{d,n}$, $f_A$ stands for the AGC frequency threshold, $P_{\Delta, max}$ represents the maximal regulation ability of the system, $H_{g,n}$ is the ramp rate of the AGC unit at bus $n$ ($H_{g,n}=0$ if no AGC unit is attached to bus $n$), and 
\begin{align}
	H_G &  = \sum\nolimits_{n=1}^N H_{g,n} \quad n \in \mathcal{S}  \notag 
\end{align}

Regarding the regulation model in \eqref{eq:FR_model}, the following should be emphasized: 1) The first segmentation in \eqref{eq:FR_model} corresponds to a sole reaction of the frequency-dependent loads at node $n$, the second segmentation additionally includes the primary frequency regulation from generators, and the third segmentation incorporates the secondary frequency regulation when only AGC units respond. 2) During the control process, both the primary and secondary controls may exist simultaneously when $|P_{\Delta}|>K_{U}f_{A}$. However, as the secondary control eliminates the frequency deviation eventually, primary frequency control resources are automatically replaced by secondary resources. Hence, from the perspective of the final steady state, only secondary frequency resources contribute. 
\subsection{Control-aware Power Flow Model}
The first step in the derivation of a control-aware power flow model is  to select a power flow model as the basis. Although the AC power flow model is the most accurate, its inherent nonlinearity raises diverse challenges when developing the desired analytical PLF method. Hence, instead of using the AC model, we use the following general form of a linear power flow model as the basis:
\begin{align}
	\boldsymbol{\varLambda} 
	\left[                 
	  \begin{array}{c}   
		\boldsymbol{\theta}_{\mathcal{S}} \\ \boldsymbol{V}_{\mathcal{L}} \\
	  \end{array}
	\right]  & = 
	\left[                 
	  \begin{array}{c}   
		\boldsymbol{P}_{\mathcal{S}} \\ \boldsymbol{Q}_{\mathcal{L}} \\
	  \end{array}
	\right] + \boldsymbol{C}
	\left[                 
	  \begin{array}{c}   
		\boldsymbol{\theta}_{\mathcal{R}} \\ \boldsymbol{V}_{\mathcal{T}} \\
	  \end{array}
	\right]   \label{eq:general_form_pf}
\end{align}
where $\boldsymbol{\theta}$ and $\boldsymbol{V}$ denote the voltage angle and voltage magnitude vectors, respectively; $\boldsymbol{P}$ and $\boldsymbol{Q}$ refer to the active and reactive power injection vectors; $\boldsymbol{\varLambda}$ and $\boldsymbol{C}$ are constant matrices composed of conductance and susceptance elements of the considered system. Besides, $\mathcal{R}$ represents the set of $V\theta$ nodes, $\mathcal{T}$ denotes the set of $PV$ and $V\theta$ nodes, and, as introduced earlier, $\mathcal{S}$ the set of $PQ$ and $PV$ nodes and $\mathcal{L}$ the set of $PQ$ buses. Note that the subscripts in \eqref{eq:general_form_pf} indicate which set the corresponding variable belongs to. For example, $\boldsymbol{\theta}_{\mathcal{S}} \in \mathfrak{R}^N $ consists of the voltage angles in set $\mathcal{S}$.





To accommodate the control contributions in \eqref{eq:FR_model} into the power flow model in \eqref{eq:general_form_pf}, we first define
\begin{align}
	& \alpha_{n, 1} = \frac{-K_{d,n}}{K_D}, \quad \alpha_{n, 2} = \frac{-K_{u,n}}{K_U}, \quad \alpha_{n, 3} = \frac{-H_{g,n}}{H_G}  \notag \\
	& \Delta_1 = 0, \ \ \Delta_2 = K_Df_D, \ \ \Delta_3 = K_Uf_A, \ \ \Delta_4 = P_{\Delta, max} \notag
\end{align}
Consequently, $P_{\Delta,n}$ can be simplified to
\begin{align}
	P_{\Delta,n} = \alpha_{n, i} P_{\Delta} \quad \text{if} \   \Delta_i < |P_{\Delta}| \leq \Delta_{i+1} \notag
\end{align}
where $i \in \{ 1,2,3\}$. After incorporating the frequency regulation, the active power injection at bus $n$ is given by:
\begin{align}
	P_n & = P_{g,n} + P_{w,n} - P_{d,n} + \alpha_{n, i} P_{\Delta} \notag \\
	& = P_{w,n} + \alpha_{n, i} \sum\nolimits_{l=1}^N P_{w,l} + \beta_{n,i} \notag
\end{align}
where 
\begin{align}
	\beta_{n,i} = P_{g,n}  - P_{d,n} + \alpha_{n, i} \sum\nolimits_{l=1}^N (P_{g,l} - P_{d,l}) \notag
\end{align}
We define 
\begin{align}
	\boldsymbol{\alpha}_i =
	\left[
	\begin{array}{cccc}
	1+\alpha_{1,i}  & \alpha_{1,i} & \cdots & \alpha_{1,i}  \\
	\alpha_{2,i}  & 1+\alpha_{2,i} & \cdots & \alpha_{2,i}  \\
	\vdots  & \vdots & \ddots & \vdots  \\
	\alpha_{N,i}  & \alpha_{N,i} & \cdots & 1+\alpha_{N,i}  \\
	\end{array}
	\right] \in \mathfrak{R}^{N\times N }   \notag
\end{align}
for $i = 1,2,3$, and further define
\begin{align}
	\boldsymbol{P}_{\mathcal{W}} & =
	\left[
	\begin{array}{ccc}
	P_{w,1} & \cdots & P_{w,N}
	\end{array}
	\right]^\top \in \mathfrak{R}^{N}   \notag \\
	\boldsymbol{Q}_{\mathcal{W}} & =
	\left[
	\begin{array}{ccc}
	Q_{w,1} & \cdots & Q_{w,M}
	\end{array}
	\right]^\top \in \mathfrak{R}^{M}   \notag \\
	\boldsymbol{\beta}_{i} & =
	\left[
	\begin{array}{ccc}
	\beta_{1,i} & \cdots & \beta_{N,i}
	\end{array}
	\right]^\top \in \mathfrak{R}^{N}   \notag \\
	\boldsymbol{\gamma} & =
	\left[
	\begin{array}{ccc}
	Q_{g,1} - Q_{d,1} & \cdots & Q_{g,M} - Q_{d,M}
	\end{array}
	\right]^\top \in \mathfrak{R}^{M}   \notag
\end{align}
Then, for $\Delta_i < |P_{\Delta}| \leq \Delta_{i+1}$, the control-aware power flow model, i.e. \eqref{eq:general_form_pf} after integrating the frequency regulation, can be summarized as 
\begin{align}
	\boldsymbol{\varLambda} 
	\left[                 
	  \begin{array}{c}   
		\boldsymbol{\theta}_{\mathcal{S}} \\ \boldsymbol{V}_{\mathcal{L}} \\
	  \end{array}
	\right]  & = 
	\boldsymbol{E}_i \boldsymbol{X} + \boldsymbol{D}_i + \boldsymbol{C}
	\left[                 
	  \begin{array}{c}   
		\boldsymbol{\theta}_{\mathcal{R}} \\ \boldsymbol{V}_{\mathcal{T}} \\
	  \end{array}
	\right]    \label{eq:pf_regulation_complex}
\end{align}
with
\begin{align}
	& \boldsymbol{X} = 
	\left[
	\begin{array}{c}
	\boldsymbol{P}_{\mathcal{W}} \\ \boldsymbol{Q}_{\mathcal{W}}
	\end{array}
	\right]  \in \mathfrak{R}^{(N+M) }, \ \  
	\boldsymbol{D}_i = 
	\left[
	\begin{array}{c}
	\boldsymbol{\beta}_{i} \\ \boldsymbol{\gamma}
	\end{array}
	\right]  \in \mathfrak{R}^{(N+M) } \notag \\
	&  \boldsymbol{E}_i = 
	\left[
	\begin{array}{cc}
	\boldsymbol{\alpha}_i & \boldsymbol{0} \\
	\boldsymbol{0} & \boldsymbol{I} \\
	\end{array}
	\right] \in \mathfrak{R}^{(N+M)\times (N+M) }  \notag
\end{align}
where $\boldsymbol{I}$ is an identity matrix with the appropriate dimension, and $\boldsymbol{0}$ is a zero matrix with the appropriate dimension.

\section{Key Challenges and Solutions}

Based on the control-aware power flow model discussed above, this section discusses the key challenges that need to be resolved when developing the desired analytical PLF method. 




\subsection{Key Challenges}
There are two key challenges that we need to address in order to develop an accurate analytical method for computing the joint PLF, i.e. the JPD of $[\boldsymbol{\theta}_{\mathcal{S}};\boldsymbol{V}_{\mathcal{L}}]$.


\textit{Challenge 1}: Normally, the historical data of the random variable $\boldsymbol{X}$, i.e. the renewable power injections, is accessible, making the JPD of $\boldsymbol{X}$ obtainable through training on the historical data. However, how to leverage this information and analytically and accurately map the JPD of $\boldsymbol{X}$ to the JPD of $[\boldsymbol{\theta}_{\mathcal{S}};\boldsymbol{V}_{\mathcal{L}}]$ given the control-aware power flow model in \eqref{eq:pf_regulation_complex} is the first key challenge that needs to be resolved.  



\textit{Challenge 2}: Assume that we addressed the first challenge and proposed a method that can realize an accurate probability distribution mapping from $\boldsymbol{X}$ to $[\boldsymbol{\theta}_{\mathcal{S}};\boldsymbol{V}_{\mathcal{L}}]$, thereby obtaining the joint PLF result. The accuracy of these results is also dependent on the precision of the adopted linear power flow model. Accordingly, how to improve the accuracy of the linear power flow model given in \eqref{eq:general_form_pf} is the second key challenge.  


\subsection{Solution to Challenge 1: Probability Distribution Mapping}
Let us simplify the power flow model in \eqref{eq:pf_regulation_complex} to
\begin{equation}
	\boldsymbol{Y} = \left\{
		\begin{split}
		& \boldsymbol{A}_1\boldsymbol{X} + \boldsymbol{b}_1 & \boldsymbol{\epsilon}^\top\!\! \boldsymbol{X} \in \boldsymbol{\Omega}_1  \\
		& \quad \quad \vdots & \vdots \quad \quad  \\
		& \boldsymbol{A}_I\boldsymbol{X} + \boldsymbol{b}_I & \boldsymbol{\epsilon}^\top\!\! \boldsymbol{X} \in \boldsymbol{\Omega}_I  \\
		\end{split} \right. \label{eq:Y_simplified}
\end{equation}
with the following auxiliary vectors and matrix
\begin{align}
	\boldsymbol{Y} & = 
	\left[                 
		\begin{array}{c}   
		\boldsymbol{\theta}_{\mathcal{S}} \\ \boldsymbol{V}_{\mathcal{L}} \\
		\end{array}
	\right]   \in \mathfrak{R}^{(N+M) } \notag \\
	\boldsymbol{\epsilon} & = \big[
		\underbrace{1 \quad \cdots \quad 1 }_{N} \quad \underbrace{0 \quad \cdots \quad 0 }_{M} \big]^\top              
	 \notag \\
	\boldsymbol{A}_i &  = \boldsymbol{\varLambda}^{-1}\boldsymbol{E}_i  \in \mathfrak{R}^{(N+M) \times (N+M) }\notag \\
	\boldsymbol{b}_i & = \boldsymbol{\varLambda}^{-1}\boldsymbol{D}_i + \boldsymbol{\varLambda}^{-1}\boldsymbol{C} 
	\left[                 
	  \begin{array}{c}   
		\boldsymbol{\theta}_{\mathcal{R}} \\ \boldsymbol{V}_{\mathcal{T}} \\
	  \end{array}
	\right] \in \mathfrak{R}^{(N+M) } \notag 
\end{align}
where $i = 1,...,I$, and $I = 3$ in the case of the proposed control-aware power flow model. Here we use $\boldsymbol{\epsilon}^\top\!\! \boldsymbol{X} \in \boldsymbol{\Omega}_i$ to equivalently represent the condition $\Delta_i < |P_{\Delta}| \leq \Delta_{i+1}$, where $\boldsymbol{\epsilon}^\top\!\! \boldsymbol{X}$ is the aggregation of active power outputs of all renewable generators. The boundaries of $\boldsymbol{\Omega}_i$ can be uniquely identified using $\Delta_{i}$, $\Delta_{i+1}$, $P_{g,n}$, and $P_{d,n}$, but we omit the explicit expressions of the boundaries of $\boldsymbol{\Omega}_i$, as they are not required in the following derivations.

After the above simplifications, the first key challenge translates into the following fundamental and general question: how to analytically derive the JPD of $\boldsymbol{Y}$ from the JPD of $\boldsymbol{X}$, given that $\boldsymbol{Y}$ is a piece-wise linear function of $\boldsymbol{X}$? To answer this question, we first use a GMM to build the JPD of $\boldsymbol{X}$ according to its historical data. Further, we prove the following proposition to pave the way for proposing the final solution.


\vspace{0.2cm}
\noindent \textbf{Proposition 1.} Let $\boldsymbol{Y}$ be a piece-wise linear function of $\boldsymbol{X}$, whose segmentations can be differentiated by the value of a linear transformation of $\boldsymbol{X}$. Then, if the JPD of $\boldsymbol{X}$ is expressed by a GMM, then the JPD of $\boldsymbol{Y}$ is an infinite GMM. 
\vspace{0.2cm}

\noindent \textit{Proof}. Since the JPD of $\boldsymbol{X}$ is a GMM, we can denote this JPD by
\begin{align}
	\mathbb{P}(\boldsymbol{X}) = \sum\nolimits_{j=1}^J w_j \ 
	\mathcal N_j(\boldsymbol{X};\boldsymbol{\mu}_j, \boldsymbol{\Sigma}_j), \quad \sum\nolimits_{j=1}^J w_j = 1 \notag 
\end{align}
where $\mathcal N_j(\cdot)$ is the $j$-th Gaussian distribution with mean $\boldsymbol{\mu}_j$, covariance $\boldsymbol{\Sigma}_j$, and weighting coefficient $w_j$. These coefficients can be trained by the expectation-maximization algorithm using the historical data of $\boldsymbol{X}$ \cite{5298967}. Then, according to \cite{7574307}, the JPD of $[\boldsymbol{X}, \boldsymbol{\epsilon}^\top\!\! \boldsymbol{X}]$ can be expressed by
\begin{align}
	\mathbb{P}([\boldsymbol{X}, \boldsymbol{\epsilon}^\top\!\! \boldsymbol{X}])\! = \! \sum\nolimits_{j=1}^J w_j \ 
	\mathcal N_j([\boldsymbol{X}, \boldsymbol{\epsilon}^\top\!\! \boldsymbol{X}];\Phi\boldsymbol{\mu}_j, \Phi\boldsymbol{\Sigma}_j\Phi^\top) \label{eq:X_joint}
\end{align}
with 
\begin{align}
	\Phi = 
	\left[                   
		\boldsymbol{I} \quad \boldsymbol{\epsilon} 
	\right]^\top \in \mathfrak{R}^{(N+M+1) \times (N+M)  } \notag
\end{align}
where $\boldsymbol{I}$ is an identity matrix with a dimension of $(N+M)$. As can be observed, $\mathbb{P}([\boldsymbol{X}, \boldsymbol{\epsilon}^\top\!\! \boldsymbol{X}])$ is still a GMM, as it is a weighted summation of $J$ Gaussian components, where the sum of the weighting coefficients is equal to 1. 

Consequently, from $\mathbb{P}([\boldsymbol{X}, \boldsymbol{\epsilon}^\top\!\! \boldsymbol{X}])$ we can analytically derive the distribution of $\boldsymbol{X}$ given the condition that $\boldsymbol{\epsilon}^\top\!\! \boldsymbol{X}=\boldsymbol{z}_l^i$ for $\forall \boldsymbol{z}_l^i \in  \boldsymbol{\Omega}_i$, where subscript $l$ indicates the $l$-th scenario in $\boldsymbol{\Omega}_i$. According to \cite{WANG2018771}, this conditional distribution can be described by 
\begin{align}
	\mathbb{P}\left[\boldsymbol{X}|\boldsymbol{\epsilon}^\top\!\! \boldsymbol{X}=\boldsymbol{z}_l^i\right] = \sum\nolimits_{j=1}^J \lambda_j \ 
	\mathcal N_j(\boldsymbol{X};\boldsymbol{\eta}_j, \boldsymbol{\Theta}_j)  \label{eq:X_conditional_first}
\end{align}
where
\begin{align}
	& \lambda_j = \frac{w_j \mathcal N_j(\boldsymbol{z}_l^i;\boldsymbol{\epsilon}^\top \boldsymbol{\mu}_{j}, \boldsymbol{\epsilon}^\top\boldsymbol{\Sigma}_j\boldsymbol{\epsilon} )}{\sum_{k=1}^J w_k 
	\mathcal N_k( \boldsymbol{z}_l^i;\boldsymbol{\epsilon}^\top \boldsymbol{\mu}_{k}, \boldsymbol{\epsilon}^\top\boldsymbol{\Sigma}_k\boldsymbol{\epsilon} )} \notag \\[2mm]
	& \boldsymbol{\eta}_j = \boldsymbol{\mu}_j + \boldsymbol{\Sigma}_j\boldsymbol{\epsilon} (\boldsymbol{\epsilon}^\top\boldsymbol{\Sigma}_j\boldsymbol{\epsilon})^{-1} (\boldsymbol{z}_l^i - \boldsymbol{\epsilon}^\top \boldsymbol{\mu}_{j})  \notag \\[2mm]
	& \boldsymbol{\Theta}_j = \boldsymbol{\Sigma}_j - \boldsymbol{\Sigma}_j\boldsymbol{\epsilon} (\boldsymbol{\epsilon}^\top\boldsymbol{\Sigma}_j\boldsymbol{\epsilon})^{-1}(\boldsymbol{\Sigma}_j\boldsymbol{\epsilon})^\top  \notag
\end{align}
Note that when $\boldsymbol{\epsilon}^\top\!\! \boldsymbol{X}=\boldsymbol{z}_l^i$, the relationship between $\boldsymbol{Y}$ and $\boldsymbol{X}$ is a pure linear mapping, i.e. $\boldsymbol{Y} = \boldsymbol{A}_i\boldsymbol{X} + \boldsymbol{b}_i$ instead of a piece-wise linear relationship. Therefore, according to the linear mapping proposition proposed in \cite{7574307}, the following holds:
\begin{equation}
	\mathbb{P}\left[\boldsymbol{Y}|\boldsymbol{\epsilon}^\top \!\! \boldsymbol{X}=\boldsymbol{z}_l^i\right] \!=\! \sum_{j=1}^J \! \lambda_j \! \ 
	\mathcal N_j(\boldsymbol{X};\boldsymbol{A}_i\boldsymbol{\eta}_j+\boldsymbol{b}_i, \boldsymbol{A}_i\boldsymbol{\Theta}_j\boldsymbol{A}_i^\top)  \label{eq:Y_conditional_first}
\end{equation}

Now, let us consider the interval $\boldsymbol{\Omega}_i$ as a complete set. By leveraging the law of total probability, we have
\begin{align}
	\mathbb{P}\! \left[\boldsymbol{Y}|\boldsymbol{\epsilon}^\top \!\! \boldsymbol{X} \! \in  \! \boldsymbol{\Omega}_i\right] = \!\!\! \sum_{\boldsymbol{z}_l^i \in  \boldsymbol{\Omega}_i} \!\!\! \mathbb{P}\! \left[\boldsymbol{\epsilon}^\top \!\! \boldsymbol{X}=\boldsymbol{z}_l^i\right] \mathbb{P} \!\left[\boldsymbol{Y}|\boldsymbol{\epsilon}^\top \!\! \boldsymbol{X}=\boldsymbol{z}_l^i\right]  \label{eq:Y_conditional}
\end{align}
where
\begin{align}
	\sum\nolimits_{\boldsymbol{z}_l^i \in  \boldsymbol{\Omega}_i} \mathbb{P}\left[\boldsymbol{\epsilon}^\top\!\! \boldsymbol{X}=\boldsymbol{z}_l^i\right] = 1 \label{eq:weighted_sum_1}
\end{align}
Since $\mathbb{P}\left[\boldsymbol{Y}|\boldsymbol{\epsilon}^\top\!\! \boldsymbol{X}=\boldsymbol{z}_l^i\right]$ is a GMM, as shown in \eqref{eq:Y_conditional_first}, then combining with \eqref{eq:weighted_sum_1}, it can be concluded that $\mathbb{P}\left[\boldsymbol{Y}|\boldsymbol{\epsilon}^\top\!\! \boldsymbol{X} \in \boldsymbol{\Omega}_i\right]$ is still a GMM, because it is still a summation of Gaussian distributions, whose weighting coefficients sum up to 1. 

We can further use the law of total probability, but this time let us consider $\boldsymbol{\Omega} = \bigcup\nolimits_{i=1}^I \boldsymbol{\Omega}_i \notag$ as the complete set. Accordingly, the JPD of $\boldsymbol{Y}$, i.e. $\mathbb{P}(\boldsymbol{Y})$, satisfies
\begin{align}
	\mathbb{P}(\boldsymbol{Y}) = \sum\nolimits_{i=1}^I \mathbb{P}\left[\boldsymbol{\epsilon}^\top\!\! \boldsymbol{X} \in \boldsymbol{\Omega}_i\right] \mathbb{P}\left[\boldsymbol{Y}|\boldsymbol{\epsilon}^\top\!\! \boldsymbol{X} \in \boldsymbol{\Omega}_i\right] \label{eq:law_total_proba}
\end{align}
Since 
\begin{align}
	\sum\nolimits_{i=1}^I \mathbb{P}\left[\boldsymbol{\epsilon}^\top\!\! \boldsymbol{X} \in \boldsymbol{\Omega}_i\right] = 1 \notag
\end{align}
holds and $\mathbb{P}\left[\boldsymbol{Y}|\boldsymbol{\epsilon}^\top\!\! \boldsymbol{X} \in \boldsymbol{\Omega}_i\right]$ is a GMM, this implies that $\mathbb{P}(\boldsymbol{Y})$ is a GMM as well, whose detailed expression is
\begin{align}
	\mathbb{P}(\boldsymbol{Y}) =  \sum_{i=1}^I\! \mathbb{P}\!\left[\boldsymbol{\epsilon}^\top \!\! \boldsymbol{X} \in \boldsymbol{\Omega}_i\right] \!\!\! \sum_{\boldsymbol{z}_l^i \in  \boldsymbol{\Omega}_i} \!\!\! \mathbb{P}\! \left[\boldsymbol{\epsilon}^\top \!\! \boldsymbol{X}=\boldsymbol{z}_l^i\right] \mathbb{P} \!\left[\boldsymbol{Y}|\boldsymbol{\epsilon}^\top \!\! \boldsymbol{X}=\boldsymbol{z}_l^i\right]   \label{eq:JPD_Y}
\end{align}

Note that $\mathbb{P}(\boldsymbol{Y})$ slightly differs from the conventional GMM, as the number of Gaussian distributions in $\mathbb{P}(\boldsymbol{Y})$ could be infinity because there might be infinite selections of $\boldsymbol{z}_l^i$ in $\boldsymbol{\Omega}_i$, and each $\boldsymbol{z}_l^i$ corresponds to a $\mathbb{P}\left[\boldsymbol{Y}|\boldsymbol{\epsilon}^\top\!\! \boldsymbol{X}=\boldsymbol{z}_l^i\right]$. Accordingly, $\mathbb{P}(\boldsymbol{Y})$ is an infinite GMM \footnote{A GMM composed of an infinite number of Gaussian components is defined as an infinite GMM \cite{rasmussen1999infinite}.}.$\hfill\blacksquare$



Based on Proposition 1, we further propose two methods, namely the direct method and the indirect method, to analytically realize the mapping from $\mathbb{P}(\boldsymbol{X})$ to $\mathbb{P}(\boldsymbol{Y})$.

\subsubsection{Direct Method} As the name indicates, this method directly uses the expression in \eqref{eq:JPD_Y} to obtain $\mathbb{P}(\boldsymbol{Y})$. However, since we are not able to aggregate infinite Gaussian components in practice, we thus use the following approximation:
\begin{align}
	\mathbb{P}(\boldsymbol{Y}) \approx  \sum_{i=1}^I\! \mathbb{P}\!\left[\boldsymbol{\epsilon}^\top \!\! \boldsymbol{X} \in \boldsymbol{\Omega}_i\right] \!\! \sum_{l=1}^L \! \mathbb{P}\! \left[\boldsymbol{\epsilon}^\top \!\! \boldsymbol{X}=\boldsymbol{z}_l^i\right] \mathbb{P} \!\left[\boldsymbol{Y}|\boldsymbol{\epsilon}^\top \!\! \boldsymbol{X}=\boldsymbol{z}_l^i\right]   \label{eq:JPD_Y_approx}
\end{align}
i.e. we limit the number of Gaussian components to $L$. Undoubtedly, different values of $L$ will affect the approximation accuracy, which will be discussed in Section V.

Summarizing: 1) $\mathbb{P}\left[\boldsymbol{Y}|\boldsymbol{\epsilon}^\top\!\! \boldsymbol{X}=\boldsymbol{z}_l^i\right]$ can be analytically obtained via \eqref{eq:Y_conditional_first}, while \eqref{eq:Y_conditional_first} is derived from $\mathbb{P}(\boldsymbol{X})$ as shown in the proof. 2) As mentioned earlier, $\boldsymbol{\epsilon}^\top\!\! \boldsymbol{X}$ is the aggregation of active power outputs of all renewable generators, thus it is straightforward to obtain $\mathbb{P}\!\left[\boldsymbol{\epsilon}^\top \!\! \boldsymbol{X} \in \boldsymbol{\Omega}_i\right]$ and $\mathbb{P}\! \left[\boldsymbol{\epsilon}^\top \!\! \boldsymbol{X}=\boldsymbol{z}_l^i\right]$ using historical or sampling data of renewable generators' output. 3) Although the direct method is straightforward to implement and is efficient as only simple algebraic calculations are required, $\mathbb{P}(\boldsymbol{Y})$ is composed of a large number of Gaussian components when $L$ is large, resulting in a high computational burden when using $\mathbb{P}(\boldsymbol{Y})$ for power system analysis. 


\subsubsection{Indirect Method} To reduce the number of Gaussian components in $\mathbb{P}(\boldsymbol{Y})$, we additionally propose an indirect method.


First, similar to \eqref{eq:Y_conditional}, $\mathbb{P}\! \left[\boldsymbol{X}|\boldsymbol{\epsilon}^\top \!\! \boldsymbol{X} \! \in  \! \boldsymbol{\Omega}_i\right]$ also satisfies
\begin{align}
	\mathbb{P}\! \left[\boldsymbol{X}|\boldsymbol{\epsilon}^\top \!\! \boldsymbol{X} \! \in  \! \boldsymbol{\Omega}_i\right] = \!\!\! \sum_{\boldsymbol{z}_l^i \in  \boldsymbol{\Omega}_i} \!\!\! \mathbb{P}\! \left[\boldsymbol{\epsilon}^\top \!\! \boldsymbol{X}=\boldsymbol{z}_l^i\right] \mathbb{P} \!\left[\boldsymbol{X}|\boldsymbol{\epsilon}^\top \!\! \boldsymbol{X}=\boldsymbol{z}_l^i\right]  \label{eq:X_conditional}
\end{align}
given the law of total probability. Recall that $\mathbb{P}\left[\boldsymbol{X}|\boldsymbol{\epsilon}^\top\!\! \boldsymbol{X}=\boldsymbol{z}_l^i\right]$ is a GMM according to \eqref{eq:X_conditional_first}. Hence, $\mathbb{P}\! \left[\boldsymbol{X}|\boldsymbol{\epsilon}^\top \!\! \boldsymbol{X} \! \in  \! \boldsymbol{\Omega}_i\right]$ is an infinite GMM since it is a weighted sum of infinite Gaussian components whose weighting coefficients add up to 1. 

However, unlike $\mathbb{P}\! \left[\boldsymbol{Y}|\boldsymbol{\epsilon}^\top \!\! \boldsymbol{X} \! \in  \! \boldsymbol{\Omega}_i\right]$ in \eqref{eq:Y_conditional}, here we can access the data of $\boldsymbol{X}$ (and the data of $\boldsymbol{\epsilon}^\top \!\! \boldsymbol{X}$), therefore it is possible to use data training to obtain the GMM form of $\mathbb{P}\! \left[\boldsymbol{X}|\boldsymbol{\epsilon}^\top \!\! \boldsymbol{X} \! \in  \! \boldsymbol{\Omega}_i\right]$, instead of summing up a large number of Gaussian components to approximate this GMM. The trained result for $\mathbb{P}\! \left[\boldsymbol{X}|\boldsymbol{\epsilon}^\top \!\! \boldsymbol{X} \! \in  \! \boldsymbol{\Omega}_i\right]$ is expressed by
\begin{equation}
	\mathbb{P}\! \left[\boldsymbol{X}|\boldsymbol{\epsilon}^\top \!\! \boldsymbol{X} \! \in  \! \boldsymbol{\Omega}_i\right] = \sum\nolimits_{j=1}^J \sigma_j \ 
	\mathcal N_j(\boldsymbol{X};\boldsymbol{\varpi}_j, \boldsymbol{\Psi}_j)  \label{eq:X_conditional_train}
\end{equation}

Given that $\boldsymbol{Y} = \boldsymbol{A}_i\boldsymbol{X} + \boldsymbol{b}_i$ when $\boldsymbol{\epsilon}^\top \!\! \boldsymbol{X} \! \in  \! \boldsymbol{\Omega}_i$, using the linear mapping proposition proposed in \cite{7574307} again leads to the analytical expression of $\mathbb{P}\! \left[\boldsymbol{Y}|\boldsymbol{\epsilon}^\top \!\! \boldsymbol{X} \! \in  \! \boldsymbol{\Omega}_i\right]$:
\begin{equation}
	\mathbb{P}\! \left[\boldsymbol{Y}|\boldsymbol{\epsilon}^\top \!\! \boldsymbol{X} \! \in  \! \boldsymbol{\Omega}_i\right] \! = \!\sum_{j=1}^J \! \sigma_j \! \ 
	\mathcal N_j(\boldsymbol{Y};\boldsymbol{A}_i\boldsymbol{\varpi}_j\!+ \! \boldsymbol{b}_i, \boldsymbol{A}_i\boldsymbol{\Psi}_j\boldsymbol{A}_i^\top)  \label{eq:Y_conditional_train}
\end{equation}
which is derived from \eqref{eq:X_conditional_train}. Once we have $\mathbb{P}\! \left[\boldsymbol{Y}|\boldsymbol{\epsilon}^\top \!\! \boldsymbol{X} \! \in  \! \boldsymbol{\Omega}_i\right]$, $\mathbb{P}(\boldsymbol{Y})$ can be easily obtained by \eqref{eq:law_total_proba}.
\begin{figure}[h]
	\centering  
	\includegraphics[width=3.5in]{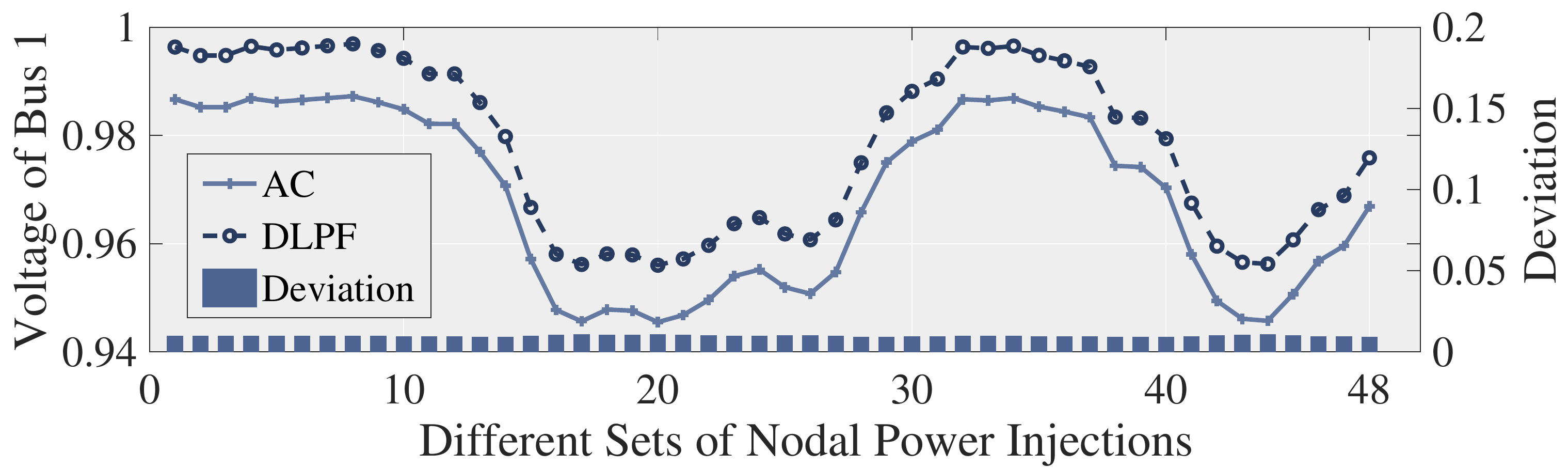} 
	\vspace{-0.7cm}
	\caption{The voltage results of bus 1 in the IEEE 118-bus system computed by the DLPF and AC models and their deviations under different sets of nodal power injections}
	\label{fig:1_constant_deviation} 
\end{figure}

Hence, the indirect method avoids summing up Gaussian components and instead uses the GMM training in \eqref{eq:X_conditional_train}. Consequently, the result of $\mathbb{P}(\boldsymbol{Y})$ only contains $I\times J$ Gaussian distributions, which is significantly less than a large number for $L$ used in the direct method, thus reducing the model complexity of $\mathbb{P}(\boldsymbol{Y})$. However, the training process (which is based on the expectation-maximization algorithm) includes iterative calculations and may increase the computational time, making the indirect method less efficient than the direct method.

\vspace{-0.2cm}
\subsection{Solution to Challenge 2: Power Flow Model Improvement}

To obtain accurate results for the joint PLF, an accurate linear power flow model is needed. Hence, here we first adopt a high-precision linear power flow model, the DLPF model \cite{7782382}, which has the same form as shown in \eqref{eq:general_form_pf} with
\begin{align}
	\boldsymbol{\varLambda} & = 
	\left[
	\begin{array}{cc}
	- \boldsymbol{B}_{\mathcal{S},\mathcal{S}}  & \boldsymbol{G}_{\mathcal{S},\mathcal{L}}  \\
	- \boldsymbol{G}_{\mathcal{L},\mathcal{S}}  & \boldsymbol{B}_{\mathcal{L},\mathcal{L}}  \\
	\end{array}
	\right] \in \mathfrak{R}^{(N+M)\times (N+M) }  \notag\\
	\boldsymbol{C} & = 
	\left[
	\begin{array}{cc}
	\boldsymbol{B}_{\mathcal{S},\mathcal{R}}^{'}  & - \boldsymbol{G}_{\mathcal{S},\mathcal{T}}  \\
	\boldsymbol{G}_{\mathcal{L},\mathcal{R}}  & -\boldsymbol{B}_{\mathcal{L},\mathcal{T}}  \\
	\end{array}
	\right] \in \mathfrak{R}^{(N+M)\times (N-M+2) }  \notag
\end{align}
where $\boldsymbol{G}$ and $\boldsymbol{B}$ refer to the conductance matrix and susceptance matrix, respectively. Note that superscript $\prime$ represents the susceptance matrix without shunt elements.

Although the performance of the DLPF model has been verified by a number of existing works \cite{8269410,8413105,9187712}, it still yields deviations from the correct AC solution. Specifically, there is an offset between the results from the DLPF and AC models which is nearly constant for each state, e.g. voltage magnitude, independent of the total power injection and how it is distributed over the grid. This phenomenon has already been shown in \cite{8289421} and discussed in \cite{9187712}. For an intuitive illustration, Fig. \ref{fig:1_constant_deviation} provides the voltage results of bus 1 in the IEEE 118-bus system, which clearly shows a nearly-constant deviation between the DLPF and AC models. Consequently, this nearly-constant deviation can be estimated by using only a limited set of pair-results of the DLPF and AC models, and then use the determined deviation to compensate for the error in the computation of other loading situations. This correction can be denoted by the following equation:
\begin{equation}
	\boldsymbol{Y}_{AC} \approx  \boldsymbol{Y}^c = \boldsymbol{Y} + \boldsymbol{\tau}_{DL} \notag
\end{equation}
where $\boldsymbol{Y}_{AC}$ represents the AC result, $\boldsymbol{\tau}_{DL}$ is the estimated constant, and $\boldsymbol{Y}^c$ is the corrected result of the DLPF model. We call this correction method the constant correction method, which indeed significantly improves the accuracy of the DLPF model according to \cite{9187712}.


While adding a constant correction is still applicable here, adding a constant value to the DLPF model is equal to shifting the PLF result, i.e. $\mathbb{P}(\boldsymbol{Y})$, to a new mean without changing its shape, i.e. the variance value of $\mathbb{P}(\boldsymbol{Y})$ remains unchanged. As the deviation is nearly-constant but not entirely constant, there is a need to also introduce a correction of the variance. Note that the variance correction is particularly important for those PLF results whose variances are quite small (e.g. the PLF of nodal voltages) and are therefore sensitive to the change in the variance value.

In fact, the nearly-constant deviation potentially implies that the results of the DLPF and AC models may obey a linear relationship. Consequently, to simultaneously adjust the mean and variance values of the PLF results for better accuracy, this paper suggests using a 1st-order polynomial fitting to realize the correction of the DLPF model. In other words, we use the following approach for the correction:
\begin{equation}
	\boldsymbol{Y}_{AC} \approx \boldsymbol{Y}^p = \boldsymbol{\rho}_i \boldsymbol{Y} + \boldsymbol{\varsigma}_i  \quad \text{if} \  \boldsymbol{\epsilon}^\top\!\! \boldsymbol{X} \in \boldsymbol{\Omega}_i \label{eq:poly_correct}
\end{equation}
where $\boldsymbol{\rho}_i \in \mathfrak{R}^{(N+M)}$ and $\boldsymbol{\varsigma}_i \in \mathfrak{R}^{(N+M)}$ are the fitting coefficients obtained from the 1st-order polynomial fitting, and $\boldsymbol{Y}^p$ is the corrected result of the DLPF model. As a result, both the mean and variance values of $\mathbb{P}(\boldsymbol{Y})$ are corrected. 

The whole correction process based on the 1st-order polynomial fitting can be divided into two phases:
\subsubsection{Phase 1, Polynomial Fitting} We randomly generate $H$ sets of nodal power injections that satisfy $\boldsymbol{\epsilon}^\top\!\! \boldsymbol{X} \in \boldsymbol{\Omega}_i~(i=1,...,3)$. Note that $H$ could be a rather small number, i.e. $<20$. Compute the power flows for these sets of nodal power injections using the DLPF and AC models, therefore obtaining $H$ pair-results for each state (e.g. nodal voltage or nodal angle) of the system. We then input the pair-results of each state to the 1st-order polynomial fitting to obtain $\boldsymbol{\rho}_i$ and $\boldsymbol{\varsigma}_i$.
\subsubsection{Phase 2, Correction} Substituting \eqref{eq:poly_correct} into \eqref{eq:Y_simplified} by replacing $\boldsymbol{Y}$ with $\boldsymbol{Y}^p$, we can therefore revise $\boldsymbol{A}_i$ and $\boldsymbol{b}_i$ using $\boldsymbol{\rho}_i$ and $\boldsymbol{\varsigma}_i$:
\begin{align}
	\boldsymbol{A}_i \leftarrow \boldsymbol{\rho}_i\boldsymbol{A}_i, \quad \boldsymbol{b}_i \leftarrow \boldsymbol{\rho}_i\boldsymbol{b}_i + \boldsymbol{\varsigma}_i \label{eq:Ab_correct}
\end{align}
Then, we can use the direct or indirect methods proposed earlier based on the corrected $\boldsymbol{A}_i$ and $\boldsymbol{b}_i$ in \eqref{eq:Ab_correct}, and therefore obtain the corrected and more accurate version of the final joint PLF result $\mathbb{P}(\boldsymbol{Y})$. 

This correction method designed above is called the polynomial-fitting correction approach in this paper, which not only can improve the precision of the DLPF model, but also further promotes the accuracy of the joint PLF result by correcting both the mean and variance values of $\mathbb{P}(\boldsymbol{Y})$. 


\section{Analytical Joint PLF Method}
Based on the solutions proposed in the previous section, this section proposes an analytical algorithm for computing the control-aware joint PLF and discusses the features of the proposed method.


Details of the proposed analytical joint PLF method are summarized in Algorithm \ref{alg:PLF_method} in the form of pseudocode. The characteristics of the proposed analytical algorithm are discussed below, with comparisons to the state-of-the-art simulation or approximated PLF methods that have taken controls into account. 

\begin{algorithm}
	\label{alg:PLF_method}
	\caption{Analytical Joint PLF Algorithm}
	\KwIn{System parameters, control parameters, $\mathbb{P}(\boldsymbol{X})$}
	\KwIn{User-set parameters, i.e. $J$, $L$, $H$ ($H \ll L$)}
    \KwOut{Joint PLF result, i.e. $\mathbb{P}(\boldsymbol{Y})$}
    Formulate the DLPF-based $\boldsymbol{A}_i$ and $\boldsymbol{b}_i$ in \eqref{eq:Y_simplified}\;
	Perform the Polynomial-fitting correction method: \\
	\quad \ \ \!\!\! Phase 1: Polynomial Fitting to get $\boldsymbol{\rho}_i$ and $\boldsymbol{\varsigma}_i$ in \eqref{eq:poly_correct}\;
	\quad \ \ \!\!\! Phase 2: Correction to revise $\boldsymbol{A}_i$ and $\boldsymbol{b}_i$ by \eqref{eq:Ab_correct}\;
	Generate $L$ samples of $\boldsymbol{X}$ from $\mathbb{P}(\boldsymbol{X})$\;
	\Switch{Method}{
            \Case{Direct Method}{
				Derive $\mathbb{P}([\boldsymbol{X}, \boldsymbol{\epsilon}^\top\!\! \boldsymbol{X}])$ in \eqref{eq:X_joint} from $\mathbb{P}(\boldsymbol{X})$\;
				Derive $\mathbb{P}\left[\boldsymbol{X}|\boldsymbol{\epsilon}^\top\!\! \boldsymbol{X}=\boldsymbol{z}_l^i\right]$ in \eqref{eq:X_conditional_first} from \eqref{eq:X_joint}\;
				Obtain $\mathbb{P}\left[\boldsymbol{Y}|\boldsymbol{\epsilon}^\top\!\! \boldsymbol{X}=\boldsymbol{z}_l^i\right]$ by \eqref{eq:Y_conditional_first}\;
				Count $\mathbb{P}\!\left[\boldsymbol{\epsilon}^\top \!\! \boldsymbol{X} \in \boldsymbol{\Omega}_i\right]$ and $\mathbb{P}\! \left[\boldsymbol{\epsilon}^\top \!\! \boldsymbol{X}=\boldsymbol{z}_l^i\right]$\;
				Obtain $\mathbb{P}(\boldsymbol{Y})$ by \eqref{eq:JPD_Y_approx}\;
			}
            \Case{Indirect Method}{
				Train $\mathbb{P}\! \left[\boldsymbol{X}|\boldsymbol{\epsilon}^\top \!\! \boldsymbol{X} \! \in  \! \boldsymbol{\Omega}_i\right]$ into \eqref{eq:X_conditional_train} with samples\;
				Obtain $\mathbb{P}\! \left[\boldsymbol{Y}|\boldsymbol{\epsilon}^\top \!\! \boldsymbol{X} \! \in  \! \boldsymbol{\Omega}_i\right]$ by \eqref{eq:Y_conditional_train}\;
				Obtain $\mathbb{P}(\boldsymbol{Y})$ by \eqref{eq:law_total_proba}\;
			}
        }
	Return $\mathbb{P}(\boldsymbol{Y})$\;	
\end{algorithm}





\subsubsection{Independent of distribution}

Due to GMM's universal adaptability, the proposed method does not need to make assumptions regarding the probability distribution of the random variable. Even if the potential distribution of the random variable is skewed, multimodal, or heavy-tailed, the proposed method is still applicable. 
All the works in \cite{zhu2014probabilistic,8845655,7764193,8582513} assume the probability distributions of random variables in advance. For example, the unscented transformation (UT) method \cite{8582513} requires the random variable to obey a symmetrical distribution. To meet this requirement, the UT method has to leverage the Nataf transformation to convert the asymmetrical random variable into the standard Gaussian scope first before computing the PLF \cite{8582513}, which significantly increases the computational burden.

\subsubsection{Easy to capture correlation}
Leveraging the covariance of GMM, the proposed method is naturally capable of characterizing the correlation between random variables. No additional operation is needed. In contrast, both the Latin-hypercube-sampling-based Monte Carlo simulation method used in \cite{8845655} and the UT method adopted in \cite{8582513} need to adopt the Gaussian Copula to establish the relationship between correlated random variables, even though the probability distributions of these variables are assumed to be known already. Note that the Gaussian Copula requires multiple integral calculations, thereby making these two methods computationally expensive. As for \cite{zhu2014probabilistic,7764193}, the correlation between random variables is ignored.



\subsubsection{Extensiveness of results}

The output of the proposed method goes beyond the marginal probability distribution of each system state --- the output is the JPD of all system states, e.g. the joint distribution of all nodal voltages. In other words, the result of the proposed method not only describes the uncertainties in system states, but also reveals the relationship between them. As a comparison, the UT method \cite{8582513} as well as the $2m+1$ method  \cite{7764193} can only generate the moment information of each system state, e.g. the mean value of a nodal voltage, and are not capable of providing the marginal probability distribution of a system state, not to mention the joint distribution of all the states. 

\subsubsection{Satisfactory efficiency}

The proposed approach has a satisfactory efficiency, because most parts of this method are analytical derivations, which only require simple algebraic calculations. Although $H$ power flow calculations are needed for the polynomial-fitting-based correction, the time spent is negligible as $H$ can be quite small. The direct method is more efficient than the indirect method. But even with the indirect method, the proposed method is still more efficient than the simulation methods or the approximated method with the Nataf transformation. Detailed comparisons are provided later.

\subsubsection{Highly accurate}

The proposed algorithm has high accuracy. Theoretically, the accuracy of the PLF method depends on two points: (1) the accuracy of the mapping between $\mathbb{P}(\boldsymbol{X})$ and $\mathbb{P}(\boldsymbol{Y})$, and (2) the precision of the power flow model used. For the first point, both the proposed direct and indirect methods are capable of analytically mapping $\mathbb{P}(\boldsymbol{X})$ to $\mathbb{P}(\boldsymbol{Y})$. Most importantly, there is no approximation in the indirect method, leading to no accuracy loss during the probability distribution mapping. Although the direct method contains an approximation in \eqref{eq:JPD_Y_approx}, as will be illustrated in the next section, this approximation can achieve a high precision when $L$ is reasonably large, thereby ensuring the accuracy of the overall probability distribution mapping. For the second point, the DLPF model we adopted already has a higher precision compared to many other linear models \cite{7782382}, with the polynomial-fitting-based correction, the precision of the DLPF model is further improved. The above two points ensure high accuracy of the proposed PLF algorithm. 

\section{Case Study}

In this section, we evaluate the proposed approaches, including the direct method, indirect method, polynomial fitting correction method, and analytical joint PLF method. 

\subsection{Setup of Case Studies}

\subsubsection{Methods for Comparison}
The methods we adopted for comparisons are the AC-based Monte Carlo (ACMC) simulation method, Nataf-transformation-based Latin hypercube sampling Monte Carlo (NLHSMC) simulation method \cite{8845655}, Nataf-transformation-based UT (NUT) method \cite{8582513}, and $2m+1$ point estimation method \cite{7764193}. 


\subsubsection{Test Systems}
For a comprehensive evaluation, the test systems used in this paper cover both small and large systems, including the IEEE 14-bus system, 39-bus New England case, 89-bus European transmission system, IEEE 118-bus system, 200-bus synthetic Illinois system, and 1354-bus European transmission system \cite{7725528}. 
These test cases are modified by introducing wind farms, whose historical active power data are from the U.S. National Renewable Energy Laboratory \cite{draxl2015overview}. The outputs of some conventional generators in these test cases are slightly adjusted to make sure that they have enough capacity to provide frequency regulation. A subset of the conventional generators in each of the test cases is assigned as the AGC units. 



\subsubsection{Parameter Settings} 
The values for the most important parameters we used for the various test cases are given in Table \ref{tab:parameters}, where $f_N$ is the nominal frequency, and the values of $K_{g,n}$ and $K_{d,n}$ are borrowed from \cite{7764193}, whose base values are either the capacity of the corresponding conventional generator or the nominal load at the corresponding bus. 
\begin{table}[h]
	\setlength{\abovecaptionskip}{0pt}
		\centering
		\renewcommand{\arraystretch}{1.4}
		\caption{The Values of Manually-set Parameters}
		\label{tab:parameters}
		\footnotesize
		\setlength{\tabcolsep}{2mm}{
		\begin{tabular}{l c c c c c c c}
		\hline
		\bfseries Parameter & $J$ & $H$ & $f_D$ &  $f_A$ & $f_N$ & $K_{g,n}$ & $K_{d,n}$ \\
		\bfseries Value & 5 & 12 & 0.01 & 0.1 & 50 & 25 p.u. & 2.6 p.u. \\ 
		\hline
	\end{tabular}}
\end{table}

Additionally, we assume that the power factor of the wind farms is 0.85 (same as in \cite{6192342}). The sample size for the ACMC simulation is 50,000, while for the NLHSMC, it is 10,000, which is the largest sample size used in \cite{8845655}.

Also, the following experiments are implemented in MATLAB 2017A and running on an i5-7267u processor with 8GB memory. The 1st-order polynomial fitting tool used for determining the correction of the DLPF model is the built-in function $\rm polyfit(\cdot)$, while the expectation-maximization algorithm used for training GMMs is the built-in function $\rm gmdistribution.fit(\cdot)$, both in MATLAB.



\subsection{Accuracy and Efficiency of Direct and Indirect Methods}
The proposed direct and indirect methods are essentially two mathematical methods that can be used for converting $\mathbb{P}(\boldsymbol{X})$ to $\mathbb{P}(\boldsymbol{Y})$, regardless of the physical meaning of the piece-wise linear relationship between $\boldsymbol{X}$ and $\boldsymbol{Y}$. Therefore, here we ignore the power system background and randomly generate a piece-wise linear relationship of the form given in \eqref{eq:Y_simplified}, where the dimensions of $\boldsymbol{Y}$ and  $\boldsymbol{X}$ are both equal to nine. Note that in this case, $\mathbb{P}(\boldsymbol{X})$ is obtained based on the historical data of nine wind farms from Maryland in the U.S. \cite{draxl2015overview}. 

\begin{figure}[h]
	\centering  
	\includegraphics[width=3in]{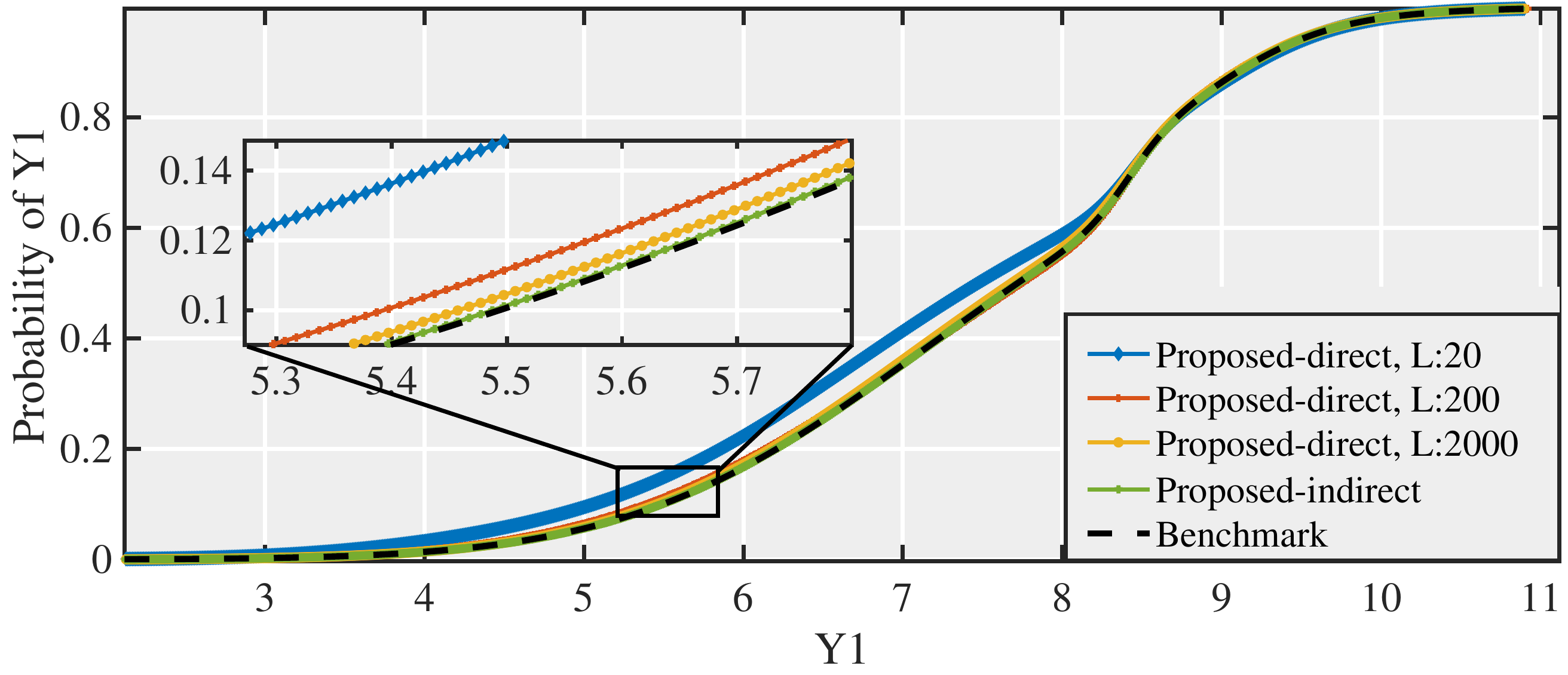} 
	\vspace{-0.3cm}
	\caption{The probability distribution mapping results obtained by the direct and indirect methods compared to the benchmark}
	\label{fig:2_direct_indirect} 
\end{figure}

For evaluation, both the direct and indirect methods are adopted to derive $\mathbb{P}(\boldsymbol{Y})$ from $\mathbb{P}(\boldsymbol{X})$, where the value of $L$ in the direct method varies from 20 to 2,000. For the benchmark of $\mathbb{P}(\boldsymbol{Y})$, we first feed 20,000 samples of $\boldsymbol{X}$ into the random piece-wise linear model and thus obtain 20,000 results of $\boldsymbol{Y}$, then we use these results to obtain a GMM with $J$ Gaussian components, which serves as the benchmark of $\mathbb{P}(\boldsymbol{Y})$ here. Note that $\mathbb{P}(\boldsymbol{Y})$ cannot be visualized directly. Therefore, we provide the marginal distribution of the first variable in $\boldsymbol{Y}$, i.e. $Y_1$, from $\mathbb{P}(\boldsymbol{Y})$ for illustration, as shown in Fig. \ref{fig:2_direct_indirect}.

Overall, except for the result of the direct method when $L=20$, the results obtained by the direct and indirect methods are close, and most importantly, the gaps between these results and the benchmark are all negligible. From the enlarged window we can conclude that: 1) as expected, the larger $L$ is, the more accurate the direct method will be. The deviation between the curves when $L=200$ and $L=2,000$ is however already small. 2) the indirect method's result coincides with the benchmark, showing a higher accuracy than the direct method's result. As mentioned earlier, this is because there is no approximation in the indirect method.


The above evaluation only includes the probability distribution of $Y_1$. To take all the other variables into account, we further use the root-mean-square error (RMSE) to measure the difference between the probability distribution of $Y_n~(n=1 \ \cdots \  9)$ obtained by each evaluated method and the corresponding benchmark. Then, we summarize the average RMSEs of the evaluated methods and list them in Table \ref{tab:RMSE_time_direct_indirect}. The RMSE comparison in Table \ref{tab:RMSE_time_direct_indirect} shows the same trend as illustrated in Fig. \ref{fig:2_direct_indirect}. Here, we just want to emphasize one point: although the accuracy of the direct method increases with an increasing $L$, when $L$ reaches 200, 
\begin{table}[h]
	\setlength{\abovecaptionskip}{0pt}
		\centering
		\renewcommand{\arraystretch}{1.4}
		\caption{The Average RMSE and Computational Time (unit: seconds) of the Direct and Indirect Method}
		\label{tab:RMSE_time_direct_indirect}
		\footnotesize
		\setlength{\tabcolsep}{1mm}{
		\begin{tabular}{l c c c c}
		\hline
		\bfseries Method & Direct,$L=20$
		& Direct,$L=200$ & Direct,$L=2,000$ &  Indirect  \\
		\bfseries RMSE & 2.38$\times 10^{-2}$ & 7.37$\times 10^{-3}$ & 4.01$\times 10^{-3}$ & 2.53$\times 10^{-3}$   \\ 
		\bfseries Time & 0.041 & 0.184 & 0.773 & 2.62   \\ 
		\hline
	\end{tabular}}
\end{table}
the accuracy of the direct method is satisfactory already and the reduction in RMSE from $L=200$ to $L=2,000$ is negligible. In other words, it does not require a large $L$ to simulate the infinite GMM. 

We also provide the computational time of the two evaluated methods in Table \ref{tab:RMSE_time_direct_indirect}. Clearly, the increase in $L$ leads to a slight rise in the calculation cost of the direct method. Besides, the direct method is more efficient than the indirect method, as no GMM training is required in the direct method. 






\subsection{Verifications on the Polynomial Fitting Correction Method}

Before moving to the evaluations of the proposed analytical joint PLF method, we would like to provide an intuitive demonstration of the performance of the polynomial-fitting-based correction. 

\begin{figure}[h]
	\centering  
	\includegraphics[width=3.5in]{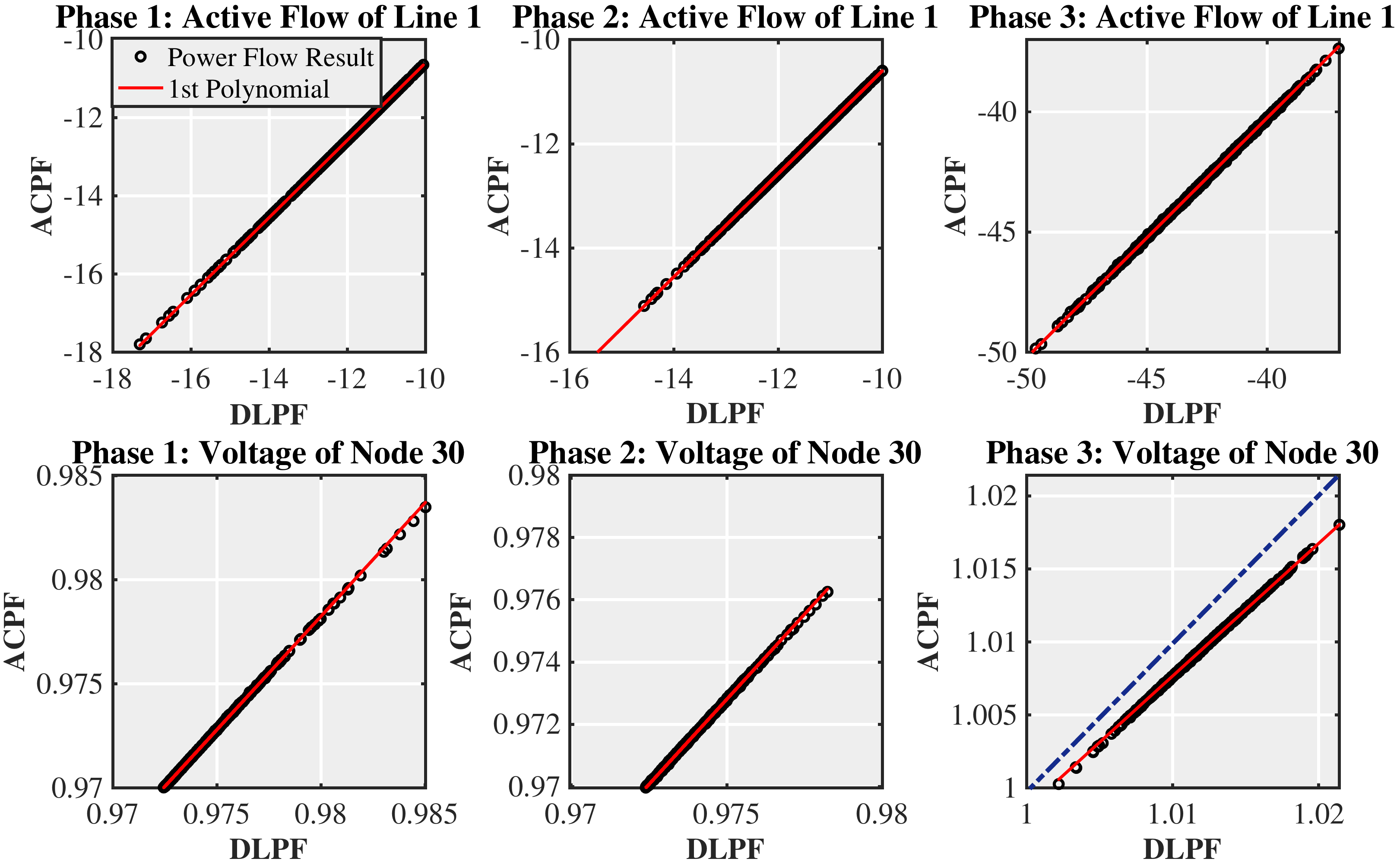} 
	\vspace{-0.5cm}
	\caption{The pair-results of the AC and DLPF models with the 1st-order polynomial fitting line for different states in the IEEE 118-bus system}
	\label{fig:3_correction} 
\end{figure}

Here, we use the IEEE 118-bus system as an example. Specifically, we first randomly generate 500 sets of nodal power injections that satisfy $\boldsymbol{\epsilon}^\top \!\! \boldsymbol{X}\in \boldsymbol{\Omega}_i ~(i=1,2,3)$. Then, we use these power injections as inputs to the AC and DLPF models, therefore obtaining 500 sets of pair-results for each state (e.g. each nodal voltage, nodal angle, or active line flow) in the system. After that, we randomly generate another 12 sets of nodal power injections that satisfy $\boldsymbol{\epsilon}^\top \!\! \boldsymbol{X} \in \boldsymbol{\Omega}_i ~(i=1,2,3)$, and use the corresponding pair-results of the AC and DLPF models as inputs to the 1st-order polynomial fitting. Both the 500 pair-results (black circle) and the polynomial fitting result (red line) are shown in Fig. \ref{fig:3_correction}, where Phase $i$ denotes the case with $\boldsymbol{\epsilon}^\top \!\! \boldsymbol{X} \in \boldsymbol{\Omega}_i ~(i=1,2,3)$. Note that due to the space limitation, in Fig. \ref{fig:3_correction}, we only illustrate the results of a few states that are visibly different for different power injections. From Fig. \ref{fig:3_correction} we can conclude two points. First, the relationship between the results obtained by the AC and DLPF models do obey a linear relationship. But most importantly, the gradients of the red lines are not always 1, as explicitly illustrated in the last sub-figure. That is to say, the assumption that the deviation between the AC and DLPF results is constant may not always hold. Second, only a small number of pair-results, e.g. 12, is sufficient to realize a satisfactory polynomial fitting.
\begin{figure}[h]
	\centering  
	\includegraphics[width=3.5in]{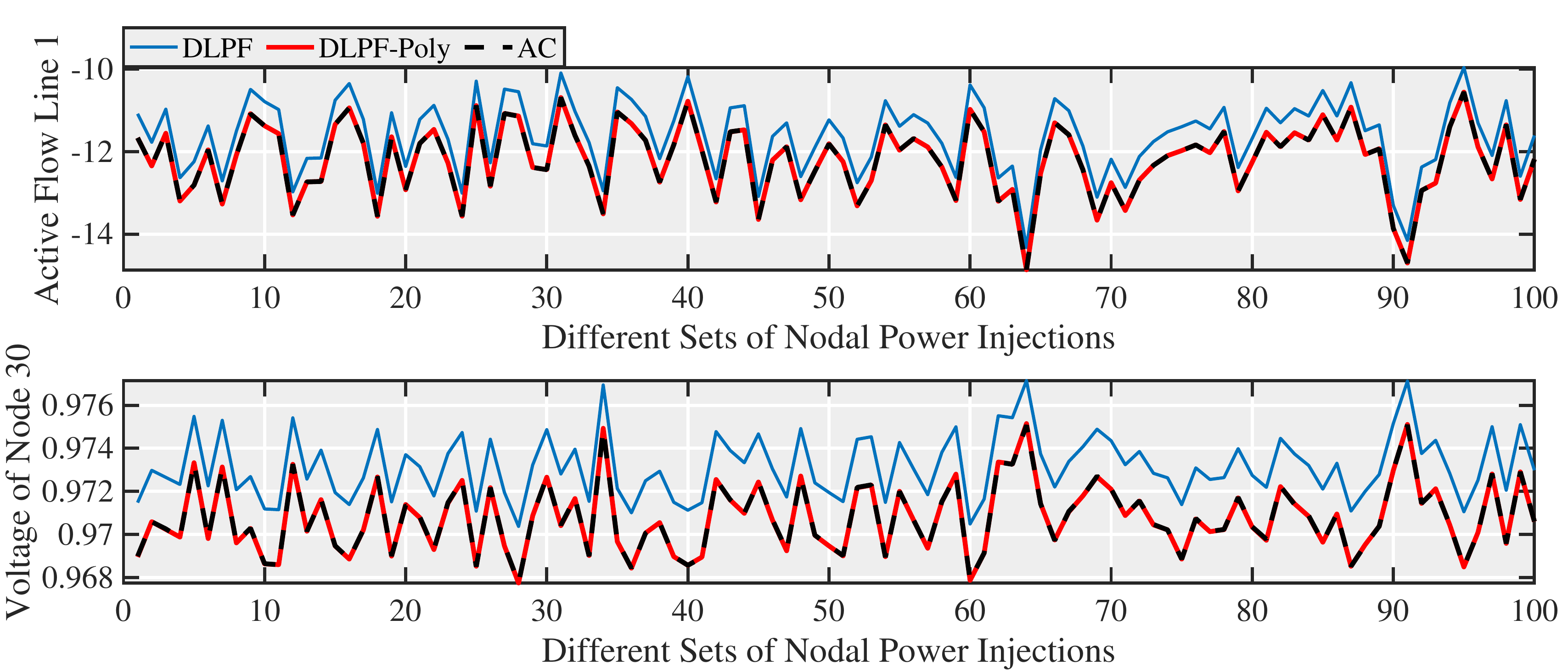} 
	\vspace{-0.5cm}
	\caption{The results of the AC model and results of the DLPF model before and after the polynomial-fitting-based correction for different states in the IEEE-118 system under different sets of nodal power injections (Poly: polynomial fitting correction method)}
	\label{fig:4_corrected_result} 
\end{figure}

After the polynomial-fitting-based correction, Fig. \ref{fig:4_corrected_result} further shows the corrected DLPF results for 100 different sets of nodal power injections, where ``DLPF-Poly'' denotes the DLPF results after the correction. It is obvious that, the corrected DLPF is far more accurate than the original DLPF. Similar results have been obtained by the other test systems.


\subsection{Method Comparison in Terms of Probability Distribution}

In this part, we verify the performance of the evaluated methods in terms of the accuracy of their probability distribution results. The methods in  the comparison include the proposed analytical joint PLF algorithm (using the indirect method), the ACMC simulation method, and the NLHSMC simulation method, where the results of the ACMC simulation method are considered as the benchmark. Note that neither the NUT nor the $2m+1$ point estimation method can generate probability distributions, hence they are excluded from the comparison here. Additionally, we also show the results for the proposed method when we use the constant-compensation-based correction instead of the polynomial fitting in the DLPF model, to highlight the performance difference between these two correction methods. For distinction, the original proposed analytical joint PLF algorithm is denoted by ``Proposed-Polynomial'' whereas the adapted version is represented by ``Proposed-Constant''. 

\begin{figure}[h]
	\centering  
	\subfigure{ 
	\includegraphics[width=3.5in]{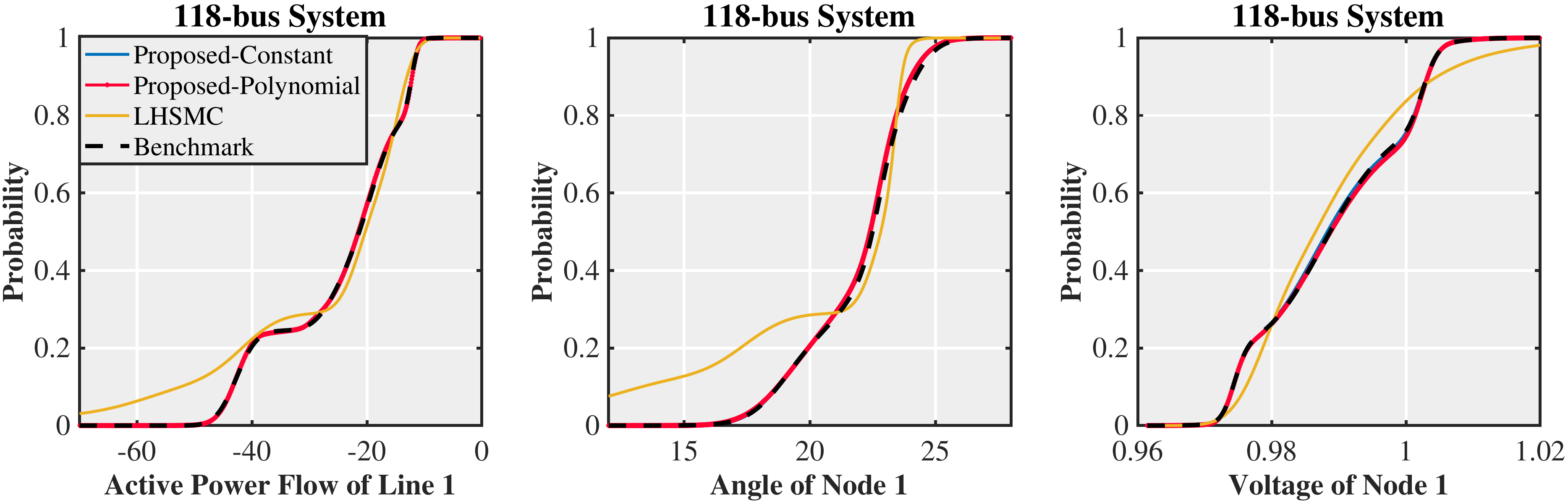} 
	} 
	\subfigure{ 
	\includegraphics[width=3.5in]{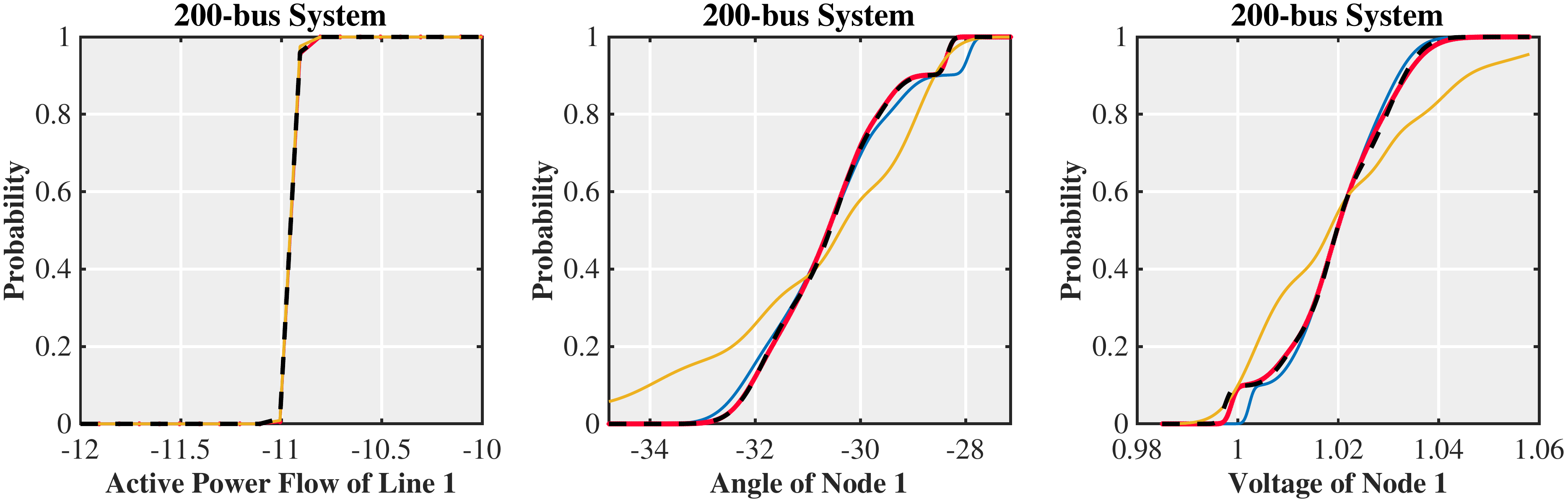} 
	}
	\vspace{-0.5cm}
	\caption{The probability distributions of different states in different test cases obtained by the evaluated methods and the benchmark}
	\label{fig:5_ACMC_CDF} 
\end{figure}

Fig. \ref{fig:5_ACMC_CDF} gives some results of the evaluated methods for different test systems. Regardless of which correction method is used, the proposed algorithm shows a high accuracy as its results perfectly coincide with the benchmark results, clearly outperforming the NLHSMC method. However, in some cases, e.g. the nodal angle and voltage in the 200-bus system, deviations occur between the proposed algorithm with the constant compensation method and the benchmark. But the proposed algorithm with the polynomial fitting correction method still shows a close-to-perfect match with the benchmark. 
\vspace{-0.3cm}
\begin{table}[h]
	\footnotesize
	\centering
	\renewcommand{\arraystretch}{1.4}
	\caption{The Average RMSE of the Evaluated Methods Compared to the ACMC Benchmark}
	\label{tab:RMSE_ACMC}
	\setlength{\tabcolsep}{1.5mm}{
	\begin{tabular}{cl>{\columncolor[gray]{0.85}}ccc}
	\hline
	\rowcolor{Whitecolor}
	\bfseries State & \bfseries System   & \bfseries Proposed-Poly      & \bfseries Proposed-Const & \bfseries NLHSMC     \\
	\hline
	\multirow{6}{*}{\textbf{Flow}}      & 14-bus   &  5.796$\times 10^{-3}$ & 6.732$\times 10^{-3}$ & 1.045$\times 10^{-1}$ \\
							   & 39-bus   & 6.811$\times 10^{-3}$ & 9.244$\times 10^{-3}$ & 5.723$\times 10^{-2}$ \\
							   & 89-bus   & 7.010$\times 10^{-3}$ & 1.356$\times 10^{-2}$ & 3.934$\times 10^{-2}$ \\
							   & 118-bus  & 4.249$\times 10^{-3}$ & 5.824$\times 10^{-3}$ & 6.727$\times 10^{-2}$ \\
							   & 200-bus  & 6.783$\times 10^{-3}$ & 1.056$\times 10^{-2}$ & 5.758$\times 10^{-2}$ \\
							   & 1354-bus & 5.306$\times 10^{-3}$ & 7.399$\times 10^{-3}$ & 8.067$\times 10^{-2}$ \\
	\hline
	\multirow{6}{*}{\textbf{Voltage}}   & 14-bus   & 8.671$\times 10^{-3}$ & 1.923$\times 10^{-2}$ & 1.046$\times 10^{-1}$ \\
							   & 39-bus   & 7.756$\times 10^{-3}$ & 3.128$\times 10^{-2}$ & 5.004$\times 10^{-2}$ \\
							   & 89-bus   & 8.445$\times 10^{-3}$ & 2.010$\times 10^{-2}$ & 3.186$\times 10^{-2}$ \\
							   & 118-bus  & 1.109$\times 10^{-2}$ & 2.985$\times 10^{-2}$ & 4.318$\times 10^{-2}$ \\
							   & 200-bus  & 8.181$\times 10^{-3}$ & 2.245$\times 10^{-2}$ & 9.097$\times 10^{-2}$ \\
							   & 1354-bus & 4.836$\times 10^{-3}$ & 2.194$\times 10^{-2}$ & 5.666$\times 10^{-2}$ \\
	\hline
	\end{tabular}}
\end{table}

To further compare the evaluated methods for all test systems and all states, we compute the average RMSEs of the results of these methods. Again, the benchmark is the result of the ACMC simulation method. Note that we ignore the states whose variances are less than $10^{-8}$, as they are basically constants without a meaningful probability distribution. The average RMSEs of the evaluated methods are listed in Table \ref{tab:RMSE_ACMC}, where the lowest values are highlighted. Clearly, the proposed algorithm with the polynomial-fitting-based correction always outperforms the other two methods. Also, the proposed algorithm has a very stable and high accuracy regardless of the scale of the test system.

\begin{table}[]
	\footnotesize
	\centering
	\renewcommand{\arraystretch}{1.4}
	\caption{The Average Relative Error of the Mean and Variance Values Obtained by the Evaluated Methods}
	\label{tab:Error_Moment}
	\setlength{\tabcolsep}{1mm}{
	\begin{tabular}{cl>{\columncolor[gray]{0.85}}cccc}
	\hline
	\multicolumn{6}{c}{ \bfseries Average Relative Error of Mean Value}                                     \\
	\hline
	\rowcolor{Whitecolor}
	\bfseries State & \bfseries System   & \bfseries Proposed & \bfseries NLHSMC   & \bfseries NUT      & \bfseries 2m+1     \\
	\hline
	\multirow{6}{*}{\textbf{Flow}}     & 14-bus   & 1.76$\times 10^{-2}$      & 1.67 & 1.80 & 1.92$\times 10^{-1}$ \\
							  & 39-bus   & 5.04$\times 10^{-4}$      & 3.47$\times 10^{-2}$ & 3.22$\times 10^{-2}$ & 5.45$\times 10^{-1}$ \\
							  & 89-bus   & 3.18$\times 10^{-3}$      & 3.40$\times 10^{-2}$ & 7.68$\times 10^{-2}$ & 7.09$\times 10^{-2}$ \\
							  & 118-bus  & 6.95$\times 10^{-3}$      & 2.17$\times 10^{-1}$ & 2.51$\times 10^{-1}$ & 1.24$\times 10^{-1}$ \\
							  & 200-bus  & 1.29$\times 10^{-2}$      & 1.97$\times 10^{-1}$ & 2.84$\times 10^{-1}$ & 4.41$\times 10^{-2}$ \\
							  & 1354-bus & 3.69$\times 10^{-4}$      & 1.59$\times 10^{-1}$ & 2.28$\times 10^{-1}$ & 1.14$\times 10^{-1}$ \\
	\hline
	\multirow{6}{*}{\textbf{Voltage}}  & 14-bus   & 6.20$\times 10^{-5}$      & 1.68$\times 10^{-3}$ & 3.27$\times 10^{-3}$ & 1.41$\times 10^{-4}$ \\
							  & 39-bus   & 7.30$\times 10^{-5}$      & 9.85$\times 10^{-4}$ & 1.09$\times 10^{-3}$ & 2.52$\times 10^{-3}$ \\
							  & 89-bus   & \cellcolor{white!00} 1.20$\times 10^{-4}$      & 2.88$\times 10^{-4}$ & 1.92$\times 10^{-4}$ & \cellcolor{gray!30}8.53$\times 10^{-5}$ \\
							  & 118-bus  & \cellcolor{white!00} 1.24$\times 10^{-4}$      & 1.29$\times 10^{-3}$ & 1.71$\times 10^{-3}$ & \cellcolor{gray!30}1.02$\times 10^{-4}$ \\
							  & 200-bus  & 5.00$\times 10^{-5}$      & 1.14$\times 10^{-3}$ & 1.25$\times 10^{-3}$ & 8.16$\times 10^{-5}$ \\
							  & 1354-bus & 2.04$\times 10^{-5}$      & 1.46$\times 10^{-3}$ & 2.08$\times 10^{-3}$ & 1.75$\times 10^{-4}$ \\
    \hline
	\hline
	\multicolumn{6}{c}{ \bfseries Average Relative Error of Variance Value}                                 \\
	\hline
	\rowcolor{Whitecolor}
	\bfseries State & \bfseries System   & \bfseries Proposed & \bfseries NLHSMC   & \bfseries NUT      & \bfseries 2m+1     \\
	\hline
	\multirow{6}{*}{\textbf{Flow}}     & 14-bus   & 6.47$\times 10^{-3}$      & 1.53 & 6.96$\times 10^{-1}$ & 1.30 \\
							  & 39-bus   & 1.40$\times 10^{-2}$      & 1.01 & 5.45$\times 10^{-1}$ & 24.3 \\
							  & 89-bus   & 1.70$\times 10^{-2}$      & 8.14$\times 10^{-1}$ & 1.93 & 5.85$\times 10^{-1}$ \\
							  & 118-bus  & 1.65$\times 10^{-2}$      & 1.74 & 9.52$\times 10^{-1}$ & 9.36$\times 10^{-1}$ \\
							  & 200-bus  & 3.19$\times 10^{-2}$      & 1.20 & 19.8 & 1.17 \\
							  & 1354-bus & 6.73$\times 10^{-3}$      & 1.80 & 2.97 & 8.59$\times 10^{-1}$ \\
	\hline
	\multirow{6}{*}{\textbf{Voltage}}  & 14-bus   & 1.83$\times 10^{-2}$      & 1.02 & 8.81$\times 10^{-1}$ & 4.84$\times 10^{-1}$ \\
							  & 39-bus   & 1.67$\times 10^{-2}$      & 8.28$\times 10^{-1}$ & 2.23$\times 10^{-1}$ & 1.05 \\
							  & 89-bus   & 9.07$\times 10^{-2}$      & 6.47$\times 10^{-1}$ & 1.22 & 6.33$\times 10^{-1}$ \\
							  & 118-bus  & 1.12$\times 10^{-1}$      & 1.87 & 8.43$\times 10^{-1}$ & 6.15$\times 10^{-1}$ \\
							  & 200-bus  & 1.41$\times 10^{-2}$      & 1.75 & 1.70 & 7.32$\times 10^{-1}$ \\
							  & 1354-bus & 3.72$\times 10^{-2}$      & 3.60 & 2.66 & 7.28$\times 10^{-1}$ \\
	\hline
	\end{tabular}}
\end{table}

\subsection{Method Comparison in Terms of Moment Information}
The performance of evaluated methods is further compared in terms of the accuracy of their moment information, i.e. the mean and variance values. The methods in the comparison include the proposed analytical joint PLF algorithm, the ACMC simulation method, the NLHSMC simulation method, the NUT method, and the $2m+1$ point estimation method. The results of the ACMC simulation method are considered as the benchmark. 

Compared to the benchmark, the average relative errors of mean and variance values for different states in each test case computed by the evaluated methods are listed in Table \ref{tab:Error_Moment}, where the lowest values are highlighted. Again, the errors of the states whose variances are less than $10^{-8}$ are ignored. The results in Table \ref{tab:Error_Moment} illustrate that, for the mean value, in most of the cases the errors of the proposed algorithm are one to three orders of magnitude smaller than the errors of other methods. In two cases, however, the error of the $2m+1$ point estimation method is the lowest, but these two lowest errors are extremely close to the corresponding errors of the proposed algorithm. Since in other cases the errors of the $2m+1$ point estimation method are significantly larger than that of the proposed algorithm, it is reasonable to conclude that the proposed method is more accurate than all the other evaluated methods in terms of the mean value. For the variance value, the proposed algorithm outperforms all other approaches. 


Additionally, the errors of the proposed method listed in Table \ref{tab:Error_Moment} are not only very small but also  independent of the test case, demonstrating the adaptability of the proposed algorithm to transmission systems of different scales. 

\subsection{Method Comparison in Terms of Computational Efficiency}
Table \ref{tab:time} summarizes the computational time of the evaluated methods for different test systems. The results show that, the proposed algorithm has a higher efficiency than the simulation methods, e.g. the NLHSMC or the ACMC methods, especially when the size of the system increases. However, the approximated method, e.g. the $2m+1$ method, requires less computational time than the proposed algorithm. But it needs to be kept in mind that the approximated methods can only generate partial information of the PLF, i.e. the moment information, whereas the proposed analytical method can acquire the same information as the simulation method, namely the JPD of system states. Additionally, the proposed algorithm is more accurate than the $2m+1$ method. Overall, the proposed method provides a very good trade-off between obtained information, accuracy and computational efficiency.
\begin{table}[]
	\footnotesize
	\centering
	\renewcommand{\arraystretch}{1.4}
	\caption{The Computational Time of the Evaluated Methods for Different Test Systems (unit:seconds)}
	\label{tab:time}
	\setlength{\tabcolsep}{1mm}{
	\begin{tabular}{lccccc}
	\hline
	\bfseries System   & \bfseries Proposed & \bfseries NLHSMC   & \bfseries NUT     & \bfseries 2m+1  & \bfseries ACMC     \\
	\hline
	14-bus   & 11.23   & 34.53   & 1.98   & 0.28 & 166.60  \\
	39-bus   & 11.70   & 52.64   & 13.33  & 0.45 & 205.15  \\
	89-bus   & 8.08    & 99.33   & 17.89  & 0.75 & 458.51  \\
	118-bus  & 12.36   & 103.09  & 41.60  & 0.71 & 307.66  \\
	200-bus  & 10.06   & 145.44  & 41.08  & 1.03 & 453.82  \\
	1354-bus & 76.93   & 1344.65 & 301.20 & 8.99 & 3298.79 \\
	\hline
	\end{tabular}}
\end{table}


\section{Conclusion}
This paper proposes an analytical, control-aware joint PLF algorithm for transmission systems; the control actions considered are the primary and secondary frequency regulations including dead-zone settings. To the best of the authors' knowledge, this is the first analytical control-aware PLF method for transmission grids. Most importantly, the proposed method can be applied to any random variables that obey arbitrary distributions, can easily capture the relationship between random variables, and can generate the JPD of all system states. Simulation results show that the proposed method exhibits acceptable computational efficiency, while providing  highly accurate results regardless of the size of the test system.

There still exist many open issues worth further investigation. For example, although the polynomial fitting correction method significantly improves the precision of the DLPF model, the principal reason for the nearly-linear relation between the results of the DLPF and AC models beckons further research. Additionally, how to extend our work to the case where the load frequency characteristic coefficient and the governor response coefficients are uncertain \cite{7764193}, is also worthy of study.

\bibliographystyle{IEEEtran}
\bibliography{IEEEabrv,paper}
\end{document}